\documentclass[pra]{revtex4}
\usepackage{hyperref}
\usepackage{amsmath}
\usepackage{amssymb}
\usepackage{amsthm}
\usepackage{graphics}
\usepackage{graphicx}
\long\def\ca#1\cb{}
\def\bra#1{\langle#1|}

\def\dya#1#2{|#1\rangle_#2\langle#1|}
\def\inpd#1#2{\langle#1|#2\rangle }
\def\ket#1{|#1\rangle }

\def\eqref#1{(\ref{#1})}
\def\Tr#1{\textrm{Tr}\left[#1\right]}

\def\EC{{\cal E}}

\def\HC{{\cal H}}
\def\OC{{\cal O}}

\def\UC{{\cal U}}

\def\BC{{\cal B}}

\newtheorem{thm1}{Theorem}
\newtheorem{cons1}{Consequence}
\newtheorem{cons2}[cons1]{Consequence}
\newtheorem{cons3}[cons1]{Consequence}
\newtheorem{cons4}[cons1]{Consequence}
\newtheorem{cons5}[cons1]{Consequence}
\newtheorem{cons6}[cons1]{Consequence}
\newtheorem{cons7}[cons1]{Consequence}
\newtheorem{cons8}[cons1]{Consequence}

\begin{document}
\title{All maximally entangling unitary gates}
\author{Scott M. Cohen$^{1,2}$}
\email{cohensm@duq.edu}
\affiliation{$^1$Department of Physics, Duquesne University, Pittsburgh,
Pennsylvania 15282\\
$^2$Department of Physics, Carnegie-Mellon University,
Pittsburgh, Pennsylvania 15213}

\begin{abstract}
We characterize all maximally entangling bipartite unitary operators, acting on systems $A,B$ of arbitrary finite dimensions $d_A\le d_{\!B}$, when use of ancillary systems by both parties is allowed. Several useful and interesting consequences of this characterization are discussed, including an understanding of why the entangling and disentangling capacities of a given (maximally entangling) unitary can differ and a proof that these capacities must be equal when $d_A=d_B$.
\end{abstract}

\date{Version of 20 February 2011}

\maketitle
\section{Introduction}
A key question in quantum information theory is to understand the communication capabilities of quantum channels, wherein information is encoded in a quantum system which is then sent through the channel, which will generally introduce noise into the state of the transmitted system. A noisy quantum channel can be modeled as a unitary interaction between the system and its environment, and it is common to assume that the environment starts out in a fixed pure state. One can, however, imagine a more general scenario, where the system and environment are allowed to have a completely arbitrary initial state. Then, we are considering the action of a unitary gate between two systems, and by varying their initial state, we can seek to maximize the amount of information communicated between the corresponding parties, where this communication may be in the form of classical information, quantum information, or both. In the case of quantum information, there is a close relationship to the amount of entanglement that can be produced by the given unitary interaction, and this is the question of interest to us here: What is the capacity of a bipartite unitary gate to generate entanglement \cite{BennettUCaps}?

We consider a unitary $\UC$ acting on systems $A$ and $B$ held by Alice and Bob, respectively, system $A$ described by Hilbert space $\HC_A$ and $B$ by $\HC_B$, these Hilbert spaces having dimensions $d_{\!A}\le d_{\!B}$. Alice and Bob are allowed the use of ancillary systems, $a$ held by Alice ($\HC_a$) and $b$ held by Bob ($\HC_b$). It is well known that the use of ancillary systems increases the capacity of a unitary to generate entanglement \cite{CiracEnt,KrausCirac}. If the input state on $AaBb$ is $|\Psi_{in}\rangle$ and the output is then $|\Psi_{out}\rangle=I_a\otimes I_b\otimes \UC|\Psi_{in}\rangle$, with $I_{a(b)}$ the identity operator on $a(b)$, the capacity to generate entanglement is defined as
\begin{align}\label{eqn0}
\EC(\UC)=\sup_{|\Psi_{in}\rangle}\left(E(\Psi_{out})-E(\Psi_{in})\right),
\end{align}
where $E(\Psi)$ measures the entanglement of $\ket{\Psi}$. The maximum possible value of $\EC(\UC)$ is $2\log d_A$, since any $\UC$ can be simulated by LOCC using this amount of entanglement as a resource (the state of Alice's system can then be teleported to Bob and back) and LOCC cannot increase entanglement \cite{BennettConcentrate}. In this paper, we will only be interested in those $\UC$ that are maximally entangling, that is, those that can increase entanglement by $2\log d_A$ ebits with some choice of $|\Psi_{in}\rangle$.

In general, it is not known how large the ancilla need be to maximize the generation of entanglement for a given $\UC$, a significant barrier to understanding the entangling capacity of unitary interactions. However, in the case that $\UC$ is maximally entangling, it has been shown that one can restrict consideration to $d_a=d_{\!A}$ and $d_b=d_{\!B}$ \cite{LSW}. There it was also shown that in this case, one may use an initial state that is product,
\begin{align}\label{eqn2}
\ket{\Psi_{in}}_{AaBb}=\ket{\Phi}_{Aa}\otimes\ket{\Psi}_{Bb},
\end{align}
with $\ket{\Phi}_{Aa}=\sum_{k=1}^{d_{\!A}}\ket{k}_a\ket{k}_A/\sqrt{d_{\!A}}$ a maximally entangled state. In the next section, we use these results to characterize all maximally entangling unitaries for any dimensions $d_A,d_B$. Then, in section~\ref{sec:consq}, we deduce several consequences of this characterization. Finally, in section~\ref{sec:conc}, we summarize what has been accomplished.

\section{Characterization of maximally entangling unitaries}\label{sec:char}
Our goal is to establish a characterization of maximally entangling bipartite unitaries. To that end, we will find it convenient to expand $\UC$, assumed to be unitary and maximally entangling, in terms of a finite group $G$, elements $f,g\in G$, group multiplication represented by $fg$. Thus, we have
\begin{align}\label{eqn3}
\UC=\sum_{f\in G}\Gamma(f)\otimes W(f),
\end{align}
where $W(f)$ act on $\HC_B$, $\{\Gamma(f)\}$ are a set of unitary matrices forming a representation of $G$ and acting on $\HC_A$, $\Gamma(f)\Gamma(g)=\mu(f,g)\Gamma(fg)$, with the quantities $\mu(f,g)$ constituting a factor system for which we have that $|\mu(f,g)|=1~\forall{f,g}$ due to the fact that $\Gamma(f)$ is unitary for each $f$. When $\mu(f,g)=1~\forall{f,g}$, we have an ordinary representation of the group; otherwise it is known as a projective representation.

We know that such an expansion is always possible, since there exist groups of order $|G|=d_{\!A}^2$ that have representations forming a complete basis of the space of $d_{\!A}\times d_{\!A}$ matrices; the generalized Pauli operators provide one such example. However, for many unitaries, smaller groups are certainly possible, so we need to address the question of how to choose $G$. In \cite{ourNLU}, we used this type of expansion of bipartite unitaries to develop protocols for implementing $\UC$ using local operations and classical communication (LOCC) with prior shared entanglement as a resource. For a given $\UC$ and any $G$ with which such an expansion of $\UC$ is possible, we showed how to deterministically simulate $\UC$ by LOCC with a resource state having entanglement equal to $\log|G|$. Since LOCC cannot increase the entanglement, we see that $\log|G|$ must be at least as large as the amount of entanglement that $\UC$ can generate. This means that for the maximally entangling unitaries we are considering here, which have the ability to generate $2\log d_{\!A}$ ebits, we need a group of order $|G|\ge d_{\!A}^2$.

Writing the (to this point unknown) initial state on $Bb$ as
\begin{align}\label{eqn4}
\ket{\Psi}_{Bb}=\sum_{m,n=1}^{d_{\!B}}M_{mn}\ket{n}_b\ket{m}_B,
\end{align}
the action of $\UC$ on the input state $\ket{\Psi_{in}}$ of \eqref{eqn2} yields
\begin{align}\label{eqn5}
\ket{\Psi_{out}}&=\sum_{f\in G}[I_a\otimes\Gamma(f)]\ket{\Phi}_{Aa}\otimes [I_b\otimes W(f)]\ket{\Psi}_{Bb}\notag\\
&=\frac{1}{\sqrt{d_{\!A}}}\sum_{j,k=1}^{d_{\!A}}\ket{k}_a\ket{j}_A\sum_{f\in G}\left[\Gamma(f)\right]_{jk}\sum_{m,n=1}^{d_{\!B}}M_{mn}\ket{n}_bW(f)\ket{m}_B\notag\\
&=\frac{1}{d_{\!A}}\sum_{j,k=1}^{d_{\!A}}\ket{k}_a\ket{j}_A\otimes\ket{b_{jk}},
\end{align}
where $\left[\Gamma(f)\right]_{jk}$ is the $jk$ matrix element of $\Gamma(f)$, this basis chosen for convenience to be that which completely reduces the $\Gamma(f)$ matrices into irreducible representations (the finest block-diagonal form of these matrices). We have defined
\begin{align}\label{eqn6}
\ket{b_{jk}}=\sqrt{d_{\!A}}\sum_{f\in G}\left[\Gamma(f)\right]_{jk}\sum_{m,n=1}^{d_{\!B}}M_{mn}\ket{n}_bW(f)\ket{m}_B.
\end{align}
Assuming that $\ket{\Psi_{in}}$ is an optimal input, achieving the maximal entanglement generation of $2\log d_{\!A}$ ebits, we see immediately from \eqref{eqn5} that the states $\ket{b_{jk}}$ must form an orthonormal set, $\delta_{jj^\prime}\delta_{kk^\prime}=\langle b_{j^\prime k^\prime}\ket{b_{jk}}$. This implies, first of all, that for each fixed $j,k$, $\exists f$ such that $[\Gamma(f)]_{jk}\ne0$. Recalling that we have chosen the $j,k$ basis to be that which completely decomposes matrices $\Gamma(f)$ into irreducible representations, we see that these matrices are themselves an irreducible representation for $G$ of dimension $d_{\!A}$. Therefore, the choice of $G$ is restricted to one which has an irreducible representation of this dimension. Since the sum of the squared dimensions of all irreducible representations of $G$ is equal to $|G|$, we here have another (related) way of seeing that $|G|\ge d_{\!A}^2$. As mentioned previously, we can always choose a representation by the generalized Pauli matrices, for which $|G|=d_{\!A}^2$ (with this choice we have a projective irreducible representation, and for the given factor system, this is the \emph{only} irreducible representation for $G$), and we will assume this choice has been made throughout the remainder of this paper. It is often convenient to define the $\Gamma(f)$ such that the factor system satisfies $\mu(e,g)=\mu(g,e)=\mu(g,g^{-1})=1,~\forall{g\in G}$. One possibility is to use $\Gamma(f)=\Gamma(m,n)=e^{i\theta_{mn}}X^mZ^n$, with $\theta_{mn}=\pi[mn~(\textrm{mod}~d_A)] /d_A$.

It is shown in appendix \ref{app1} that as a consequence of Schur's orthogonality relations for group representations \cite{Schensted}, the orthonormality condition on states $\ket{b_{jk}}$ is equivalent to a corresponding orthonormality condition on operators $W(f)$,
\begin{align}\label{eqn7}
\Tr{W(f)MM^\dag W(g)^\dag}=\frac{1}{d_{\!A}^2}\delta(f,g),~\forall{f,g\in G},
\end{align}
where $\delta(f,g)=1$ when $f=g$, and otherwise is equal to zero. Thus we have our main result:
\begin{thm1}\label{thm1}The bipartite unitary $\UC$ is maximally entangling iff there exists a positive semi-definite `metric' $MM^\dag$ such that \eqref{eqn7} is satisfied $\forall{f,g\in G}$, where operators $W(f)$ are obtained from an expansion of $\UC$ as in \eqref{eqn3}, with the $\Gamma(f)$ taken to be the generalized Pauli operators. The operator $M$ defines an optimal input state on systems $bB$ through \eqref{eqn4}.
\end{thm1}
In the next section, we discuss consequences of this result.

\section{Consequences of Theorem \ref{thm1}}\label{sec:consq}
\begin{cons1}\label{cons1}
Method to check if $\UC$ is maximally entangling.
\end{cons1}
\noindent Given bipartite unitary $\UC$, theorem \ref{thm1} provides a method of determining whether or not $\UC$ is maximally entangling. One need only expand $\UC$ in terms of the generalized Pauli operators, identify the set of operators $\{W(f)\}$, and then check to see if there exists a positive semi-definite operator to play the role of $MM^\dag$ such that \eqref{eqn7} is satisfied. One way to do this is to form all products, $W(g)^\dag W(f),~f\ne g$, reshape each into a column vector (such as by stacking individual columns of each product one on top of the other) and collect all these columns into a matrix. The nullspace of this matrix corresponds (by reshaping vectors in this nullspace back into matrices) to the space of all operators orthogonal to the $W(g)^\dag W(f),~f\ne g$, as is required to satisfy \eqref{eqn7}. One then needs to search, perhaps numerically, for positive operators in this nullspace. This is relatively easy to do, at least for small enough nullspaces.

\begin{cons2}\label{cons2}
Design of maximally entangling unitaries.
\end{cons2}
\noindent Theorem \ref{thm1} also allows one to design unitaries that are maximally entangling. This amounts to choosing operator $M$ and a set of $d_A^2$ linearly independent operators $W(f)$ that satisfy \eqref{eqn7}. In addition, there is also the necessity that the chosen set of $W(f)$ are such that $\UC$ is unitary. When the dimensions are not too large, it is straightforward and reasonably fast to numerically generate a maximally entangling unitary in this way (for $d_A=4,~d_B=8$ it takes less than $10$ minutes on my laptop).

\begin{cons3}\label{cons3}
Characterizing maximally entangling interaction Hamiltonians for two-qubit systems.
\end{cons3}
\noindent A characterization of two-qubit maximally entangling Hamiltonians $H$ has been given in \cite{H2x2}. Using the well-known result \cite{CiracEnt} that up to local unitaries, every two-qubit unitary may be written as $\UC=e^{-iH}$ (the usual factor $t/\hbar$ is here absorbed into the definition of $H$ for notational convenience) with
\begin{align}\label{eqn100}
H=\sum_{j=x,y,z}\alpha_j\sigma_j\otimes\sigma_j,
\end{align}
they showed that for $\UC$ to be maximally entangling, it must be that $\cos^2\alpha_x=1/2=\cos^2\alpha_y$, with the value of $\alpha_z$ being unconstrained (permutations of $\{x,y,z\}$ are also allowed, of course). In appendix \ref{app2}, we provide an alternative proof of this result based on \eqref{eqn7}. This means there is a continuum of maximally entangling two-qubit unitaries ranging from the double CNOT ($\alpha_z=0$) to the SWAP ($\cos^2\alpha_z=1/2$).

\begin{cons4}\label{cons4}
Operators $W(f)$ must form a linearly independent set.
\end{cons4}
\noindent This is easily proven, as is shown at the end of appendix \ref{app1}. Notice also that by theorem 4 of \cite{ourNLU} and for whatever group $G$ and representation $\Gamma$ are chosen for the expansion of maximally entangling $\UC$, the number of linearly independent operators in the collection $\{\Gamma(f)\}$ is $d_{\!A}^2$, because only the single $d_{\!A}$-dimensional irreducible representation appears in these matrices. This is consistent with the fact that the Schmidt rank of $\UC$ must be at least as large as the ratio of the Schmidt rank of the output state to that of the input state. That is, since our input state has Schmidt rank of one and the output state has Schmidt rank of $d_{\!A}^2$, $\UC$ must have Schmidt rank of $d_{\!A}^2$ as well.

\begin{cons5}\label{cons5}
Input state on $Bb$ is uniquely determined up to local unitaries when $d_A=d_B$, and must be a maximally entangled state.
\end{cons5}
\noindent This was proven in \cite{LSW}; we provide an alternative proof based on \eqref{eqn7} in appendix \ref{app1}.

\begin{cons6}\label{cons6}
Why the entangling and disentangling powers can be unequal.
\end{cons6}
\noindent It is now easy to see for a maximally entangling unitary how the entangling and disentangling powers can be unequal \cite{LSW}. Recall that the disentangling power of $\UC$ is just the entangling power of $\UC^\dag$. Therefore for the disentangling power, we must replace the set $\{W(f)\}$ by $\{W(f)^\dag\}$ in \eqref{eqn7}. Then, for $\UC$ to be maximally disentangling, we require the existence of an $M^\prime M^{\prime\dag}$ orthogonal to the set of operators $\{W(g)W(f)^\dag\},~\forall{f\ne g\in G}$, whereas for maximally entangling, the orthogonality requirement applies to the generally different set, $\{W(g)^\dag W(f)\},~\forall{f\ne g\in G}$. In addition, there is the normalization condition for $f=g$, and this again applies to a generally different set of operators in the two cases. As an example, \cite{LSW} provided the original demonstration that the entangling and disentangling powers can be unequal by constructing a specific maximally entangling $\UC$ and then showing that $\UC^\dag$ has strictly less than the maximum entangling power. We have calculated the $W(f)$ for their $\UC$ and find that it is easy to satisfy \eqref{eqn7} with these $W(f)$ (set $MM^\dag=[\dya{1}{B}+\dya{3}{B}]/2$), but find (numerically) that it is not possible to do so when the set $\{W(f)\}$ is replaced by $\{W(f)^\dag\}$ (one choice that almost works is to set $M^\prime M^{\prime\dag}=c_0\dya{1}{B}+c(\dya{2}{B}+\dya{3}{B})$, which satisfies orthogonality, but the normalizations cannot all be the same no matter how $c_0,c$ are chosen).

\begin{cons7}\label{cons7}
Entangling and disentangling powers are equal for maximally entangling unitaries on $d\times d$ systems.
\end{cons7}
\noindent It was shown in \cite{BerrySanders03a} that the entangling and disentangling powers of any $\UC$ are equal when $d_A=2=d_B$. We can now extend this result to arbitrary dimensions $d_A=d_B$ when restricting to maximally entangling unitaries. From consequence \ref{cons5}, we have that $MM^\dag$ must be proportional to $I_B$. Therefore, a replacement $\{W(f)\}\rightarrow \{W(f)^\dag\}$ makes no difference whatsoever in \eqref{eqn7}, from which this claim follows immediately. That is, when $d_A=d_B$ and $\UC$ is maximally entangling, then $\UC^\dag$ is also maximally entangling.

\begin{cons8}\label{cons8}
If $d_B$ is large enough compared to $d_A$, it can be that no ancillary system is needed on Bob's side.
\end{cons8}
\noindent We here provide a construction of operators $W(f)$ corresponding to $\UC$ for which system $b$ is not needed. This requires only that the first columns of the different $W(f)$ operators are mutually orthogonal and have norm equal to $1/d_A$ (the remaining part of each $W(f)$ is unconstrained apart from the requirement that $\UC$ is unitary). Then we have that the matrix element $\bra{1}W(g)^\dag W(f)\ket{1}=\delta(f,g)/d_A^2$. Choosing $MM^\dag=\dya{1}{B}$ shows that \eqref{eqn7} is satisfied $\forall{f,g}$. This choice of $MM^\dag$ corresponds to a product state across $B/b$, so system $b$ never plays a role and may be discarded. Recalling that there are $d_A^2$ different $W(f)$ operators, the mutual orthogonality of their first columns is possible only when the length $d_B$ of those columns is at least $d_A^2$. Hence, this construction is only possible when $d_B\ge d_A^2$. Then there is a $d_A^2$-dimensional subspace of $\HC_B$ that becomes maximally entangled with systems $Aa$, the remaining space not being involved in the process. Thus, it is almost as if system $B$ has the ancillary system already embedded within itself, which is most clearly understood when $d_B=d_A^2=d_A\times d_A$. In this case, $B$ can be thought of as itself consisting of two $d_A$-dimensional systems, one of which plays the role of ancillary $b$.

\section{Conclusions}\label{sec:conc}
We have given a characterization of all maximally entangling bipartite unitaries for any dimensions $d_A\le d_B$. This allows one to check if a given unitary is maximally entangling, to construct maximally entangling unitaries, and to determine optimal input states that achieve the maximal generation of entanglement. It also provides an understanding of why the entangling and disentangling capacities can differ, as well as a proof that this can only happen when $d_B>d_A$. We also saw that for $d_B\ge d_A^2$, it is possible that no ancillary system is needed on Bob's side. Finally, we have given an alternative method of characterizing maximally entangling Hamiltonians for two-qubit systems \cite{H2x2}. An interesting open question is to determine what Hamiltonians can be maximally entangling in higher-dimensional systems.

\section{Acknowledgments}
This work has been supported in part by the National Science Foundation through Grant PHY-0757251, as well as by a grant from the Research Corporation.

\appendix
\section{Proof of theorem~\ref{thm1}}\label{app1}
Here we show that orthonormality of the states $\ket{b_{jk}}$ defined in \eqref{eqn6} is equivalent to condition \eqref{eqn7} on operators $W(f)$, which appear in an expansion of $\UC$ of the form \eqref{eqn3} with $|G|=d_A^2$. From \eqref{eqn6}, we have
\begin{align}\label{eqnApp1}
	\inpd{b_{j^\prime k^\prime}}{b_{jk}}&=d_{\!A}\sum_{f,g\in G}\left[\Gamma(g)\right]_{j^\prime k^\prime}^\ast\left[\Gamma(f)\right]_{jk}\sum_{m,n=1}^{d_{\!B}}\sum_{m^\prime,n^\prime=1}^{d_{\!B}}M_{m^\prime n^\prime}^\ast M_{mn}\inpd{n^\prime}{n}\bra{m^\prime}W(g)^\dag W(f)\ket{m}\notag\\
	&=d_{\!A}\sum_{f,g\in G}\left[\Gamma(g)\right]_{j^\prime k^\prime}^\ast\left[\Gamma(f)\right]_{jk}\sum_{m^\prime,m=1}^{d_{\!B}}\left[MM^\dag\right]_{mm^\prime}\bra{m^\prime}W(g)^\dag W(f)\ket{m}\notag\\
	&=d_{\!A}\sum_{f,g\in G}\left[\Gamma(g)\right]_{j^\prime k^\prime}^\ast\left[\Gamma(f)\right]_{jk}\Tr{MM^\dag W(g)^\dag W(f)}.
\end{align}
First notice that if $\Tr{W(f)MM^\dag W(g)^\dag}=\delta(f,g)/d_{\!A}^2$, the right-hand side of this equation becomes $\sum_f\left[\Gamma(f)\right]_{j^\prime k^\prime}^\ast\left[\Gamma(f)\right]_{jk}/d_{\!A}$. However, considering the $d_{\!A}^2$ vectors $\vec\gamma_{jk},~j,k=1,\ldots,d_{\!A},$ whose components (labeled by $f\in G$) are given by
\begin{align}\label{eqnApp2}
(\vec\gamma_{jk})_f=\frac{1}{\sqrt{d_{\!A}}}\left[\Gamma(f)\right]_{jk},
\end{align}
then by Schur's orthogonality relations for irreducible representations \cite{Schensted} and the fact that the $\Gamma(f)$ representation is irreducible, these vectors form a complete orthonormal basis for the $d_{\!A}^2$-dimensional space in which they lie (recall that $|G|=d_A^2$ is the dimension of these vectors). That is, 
\begin{align}\label{eqnApp3}
\sum_f\left[\Gamma(f)\right]_{j^\prime k^\prime}^\ast\left[\Gamma(f)\right]_{jk}/d_{\!A}=\delta_{jj^\prime}\delta_{kk^\prime},
\end{align}
which yields one of the implications we sought to prove.

To prove the converse, define $d_{\!A}^2\times d_{\!A}^2$ matrix $\OC$, with matrix elements labeled by $f,g\in G$ given by
\begin{align}\label{eqnApp4}
[\OC]_{gf}=\Tr{W(f)MM^\dag W(g)^\dag}.
\end{align}
Then if $\inpd{b_{j^\prime k^\prime}}{b_{jk}}=\delta_{jj^\prime}\delta_{kk^\prime}$, \eqref{eqnApp1} can be written as
\begin{align}\label{eqnApp5}
\frac{1}{d_{\!A}^2}\delta_{jj^\prime}\delta_{kk^\prime}=\vec\gamma_{j^\prime k^\prime}^\dag\cdot\OC\cdot\vec\gamma_{jk}.
\end{align}
By \eqref{eqnApp5}, $\OC\cdot\vec\gamma_{jk}$ is orthogonal to every vector in the complete basis of the $\vec\gamma$-vectors except for one, that being $\vec\gamma_{jk}$. Therefore, $\forall{j,k}$, $\OC\cdot\vec\gamma_{jk}$ is proportional to $\vec\gamma_{jk}$, and the proportionality constant is equal to $1/d_{\!A}^2$, independent of $j,k$, again by \eqref{eqnApp5}. Thus, we have that $\OC= I/d_{\!A}^2$, where $I$ is the $d_{\!A}^2\times d_{\!A}^2$ identity matrix. Finally, recalling the definition of $\OC$ in \eqref{eqnApp4}, we have
\begin{align}\label{eqnApp6}
\frac{1}{d_{\!A}^2}\delta(f,g)=\Tr{W(f)MM^\dag W(g)^\dag},
\end{align}
which completes the proof.\hspace{\stretch{1}}$\blacksquare$

A necessary condition for \eqref{eqnApp6} to be satisfied is that the collection of $|G|=d_A^2$ operators $W(f)$ are linearly independent. This is easily seen by contradiction, so assume they are linearly dependent. Then,
\begin{align}\label{eqnApp7}
0=\sum_{f\in G}c(f)W(f),
\end{align}
for some coefficients $c(f)$ not all equal to $0$. Multiply this expression by $MM^\dag W(g)^\dag$ for each fixed $g\in G$ and then take the trace to obtain from \eqref{eqnApp6} that
\begin{align}\label{eqnApp8}
0=\sum_{f\in G}c(f)\Tr{W(f)MM^\dag W(g)^\dag}=\frac{c(g)}{d_A^2},
\end{align}
assuming \eqref{eqnApp6}. This says that $c(g)=0~\forall{g\in G}$, which contradicts the assumption of linear dependence and proves the claim.

We now give an alternate proof (see also \cite{LSW}) that $\rho=MM^\dag$ is uniquely determined when $\UC$ is maximally entangling and $d_A=d_B$. Indeed, by contradiction, assume both $\rho$ and $\rho^\prime$ serve our purpose. Then from \eqref{eqnApp6},
\begin{align}\label{eqnApp9}
0=\Tr{W(f)(\rho-\rho^\prime)W(g)^\dag}~\forall{f,g\in G},
\end{align}
which must hold even when $f=g$. This says that for each $f,g\in G$, $W(f)(\rho-\rho^\prime)$ is orthogonal to $W(g)$. However, as we have just seen, the $d_A^2=d_B^2$ operators $W(g)$ are linearly independent, hence span the entire space $\BC(\HC_B)$ of operators acting on $\HC_B$. Therefore, it must be that
\begin{align}\label{eqnApp10}
W(f)(\rho-\rho^\prime)=0
\end{align}
for every $f\in G$. Now, choose coefficients $e(f)$ such that $I_B=\sum_fe(f)W(f)$, which can always be done since $W(f)$ are a basis of $\BC(\HC_B)$. Multiplying \eqref{eqnApp10} by $e(f)$ and summing over $f$ we obtain $0=\rho-\rho^\prime$, proving the claim.

\section{Two-qubit maximally entangling Hamiltonians}\label{app2}
Using \eqref{eqn100} gives $\UC=e^{-iH}=\sum_{f}k_f\sigma_f\otimes\sigma_f$ with $f=e,x,y,z$ labeling the group element ($e$ is the identity element). From this we identify
$W_f=k_f\sigma_f$ ($\sigma_e=I$, the two-by-two identity matrix) , where
\begin{align}\label{eqnApp11}
k_e=c_xc_yc_z-s_xs_ys_z,\notag\\
k_x=c_xs_ys_z-s_xc_yc_z,\notag\\
k_y=s_xc_ys_z-c_xs_yc_z,\notag\\
k_z=s_xs_yc_z-c_xc_ys_z,
\end{align}
and we've used the abbreviations $c_f=\cos\alpha_f$ and $s_f=\sin\alpha_f$. Applying the condition \eqref{eqn7} with $MM^\dag=I/2$ (because $d_A=d_B$), the orthogonality conditions ($f\ne g$) are automatically satisfied because the Pauli operators are themselves mutually orthogonal. Therefore, we only need to worry about normalizations ($f=g$ in \eqref{eqn7}), which give
\begin{align}\label{eqnApp12}
c_x^2c_y^2c_z^2+s_x^2s_y^2s_z^2=1/4,\notag\\
c_x^2s_y^2s_z^2+s_x^2c_y^2c_z^2=1/4,\notag\\
s_x^2c_y^2s_z^2+c_x^2s_y^2c_z^2=1/4,\notag\\
s_x^2s_y^2c_z^2+c_x^2c_y^2s_z^2=1/4.
\end{align}
It not too difficult to show that these lead to the necessary and sufficient condition that two of the $\alpha$'s must have squared cosines equal to $1/2$, the third $\alpha$ being unconstrained, which is what we set out to prove.


\end{document}